\documentclass{article}
\usepackage{spconf,amsmath,graphicx,hyperref}
\usepackage{txfonts}
\usepackage{enumitem}
\usepackage{color,xcolor}
\usepackage{colortbl}
\usepackage{threeparttable}
\usepackage{subfigure}
\usepackage{array}
\usepackage{amsmath}
\newcommand{\PreserveBackslash}[1]{\let\temp=\\#1\let\\=\temp}
\newcolumntype{C}[1]{>{\PreserveBackslash\centering}p{#1}}
\newcolumntype{R}[1]{>{\PreserveBackslash\raggedleft}p{#1}}
\newcolumntype{L}[1]{>{\PreserveBackslash\raggedright}p{#1}}

\title{A Stabilized Hybrid Active Noise Control Algorithm of GFANC and FxNLMS with Online Clustering}

\name{Zhengding Luo$^{1}$, Haozhe Ma$^{2}$\textsuperscript{*}, 
Boxiang Wang$^{1}$, Ziyi Yang$^{1}$, Dongyuan Shi$^{3}$, 
Woon-Seng Gan$^{1}$\thanks{*Corresponding author: Haozhe Ma. 
The code will be available at \href{https://github.com/Luo-Zhengding/GFANC-FxNLMS}{https://github.com/Luo-Zhengding/GFANC-FxNLMS}.}
}

\address{$^{1}$School of Electrical \& Electronic Engineering, 
Nanyang Technological University, Singapore \\
$^{2}$School of Computing, National University of Singapore, Singapore \\
$^{3}$Northwestern Polytechnical University, China}

\begin{document}
\maketitle


\begin{abstract}
The Filtered-x Normalized Least Mean Square (FxNLMS) algorithm suffers from slow convergence and a risk of divergence, although it can achieve low steady-state errors after sufficient adaptation. In contrast, the Generative Fixed-Filter Active Noise Control (GFANC) method offers fast response speed, but its lack of adaptability may lead to large steady-state errors. This paper proposes a hybrid GFANC–FxNLMS algorithm to leverage the complementary advantages of both approaches. In the hybrid GFANC–FxNLMS algorithm, GFANC provides a frame-level control filter as an initialization for FxNLMS, while FxNLMS performs continuous adaptation at the sampling rate. Small variations in the GFANC-generated filter may repeatedly reinitialize FxNLMS, interrupting its adaptation process and destabilizing the system. An online clustering module is introduced to avoid unnecessary re-initializations and improve system stability. Simulation results show that the proposed algorithm achieves fast response, very low steady-state error, and high stability, requiring only one pre-trained broadband filter.
\end{abstract}

\vspace*{-0.1cm}
\vspace*{-0.4cm}

\section{Introduction}\vspace*{-0.3cm}
With the rapid advancement of technology and industry, acoustic noise problems have become increasingly severe. It is well recognized that common noise in daily life, especially in vehicle cabins, aircraft cabins, and household appliances, is predominantly low-frequency \cite{Understand-ANC}. For low-frequency noises, passive methods (e.g., earmuffs) are either ineffective or tend to be costly and bulky \cite{kuo2006active,HuangGongping}. To address this challenge, Active Noise Control (ANC) has been extensively developed \cite{Elliott,Simon}. The fundamental principle of ANC is the destructive interference of acoustic waves, first proposed by Lueg in 1936 \cite{LuegPatent}. Advances in signal processing algorithms and electronics have enabled many successful ANC implementations over the past two decades \cite{ZhangJihui,VaryPeter}.

Traditional ANC systems typically employ adaptive algorithms, such as the Filtered-x Least Mean Square (FxNLMS), to estimate the optimal coefficients of the control filter \cite{NLMS,QiuXiaojun}. However, these algorithms generally rely on the feedback error signal to update the control filter coefficients and use zero initialization for the filter \cite{ChenJingdong,Zhaosipei}. As a result, lengthy convergence is often required, and noticeable noise reduction is not achieved immediately \cite{Marek}. Moreover, adaptive ANC algorithms are highly susceptible to instability and divergence due to several practical imperfections \cite{BoxiangICASSP,YangjunSFANC,Kajikawa-SFANC}.

\vspace*{-0.1cm}
To achieve rapid response and high stability, a Selective Fixed-filter ANC (SFANC) method was proposed in \cite{DYSFANC-CNN,luoSFANCMssp}. In SFANC, a Convolutional Neural Network (CNN) automatically selects suitable pre-trained control filters for different primary noises. However, SFANC relies on a fixed set of pre-trained control filters, which can lead to degraded performance when the encountered noises differ significantly from those used during filter training \cite{SFANC-FxNLMSsubjective,Kajikawa-SFANC-Adaptive}. To overcome this limitation, the Generative Fixed-filter ANC (GFANC) method was introduced \cite{LuoGFANCBayes}. The GFANC method generates various control filters by combining sub control filters, decomposed from a pre-trained broadband filter, using weight vectors predicted by a CNN \cite{LuoGFANCMSSP}. With flexible filter generation, GFANC not only improves noise reduction levels but also significantly reduces the filter-training efforts compared with SFANC.

\begin{table*}[tp]
\caption{Comparison of the proposed GFANC-FxNLMS method with related ANC methods.}
\centering
\resizebox{0.9\linewidth}{!}{
\begin{tabular}{m{2.75cm}|m{5.2cm}|m{1.3cm}|m{1.8cm}|m{2.7cm}|m{2.6cm}}
\hline
\textbf{Method} & \textbf{Control filter update} & \textbf{Response} & \textbf{Adaptability} & \textbf{Steady-state error} & \textbf{Pre-trained filter}\\
\hline
GFANC & Filter generated by CNN & Fast & Low & Possibly high & Single filter\\
SFANC & Filter selected by CNN & Fast & Low & Possibly high & Multiple filters\\
FxNLMS & Filter updated using error signal & Slow & High & Low & None\\
SFANC-FxNLMS & FxNLMS updates the selected filter & Fast & High & Low & Multiple filters\\
\textbf{GFANC-FxNLMS} & FxNLMS updates the generated filter & Fast & High & Lower & Single filter\\
\hline
\end{tabular}
}
\label{Table Compare GFANC SFANC FxLMS}
\end{table*}

\vspace*{-0.1cm}
Conventional adaptive ANC algorithms, such as FxNLMS, can achieve low steady-state errors after sufficient adaptation. SFANC and GFANC methods provide fast response and are free from the risk of divergence, but their lack of adaptability can result in large steady-state errors. A hybrid SFANC-FxNLMS algorithm \cite{SFANC-FxNLMSLuo} was proposed to combine the advantages of SFANC and FxNLMS. While prior studies \cite{LuoGFANCBayes,LuoGFANCMSSP} have shown the superiority of GFANC over SFANC, the integration of GFANC with FxNLMS has not been explored to surpass SFANC-FxNLMS. However, directly combining GFANC and FxNLMS introduces instability due to minor weight vector variations, highlighting the necessity of a stabilized hybrid GFANC–FxNLMS algorithm.

\vspace*{-0.1cm}
To effectively leverage the strengths of GFANC and FxNLMS, this paper proposes a hybrid GFANC–FxNLMS algorithm with online clustering. The comparison of GFANC\-FxNLMS with related ANC algorithms is summarized in Table~\ref{Table Compare GFANC SFANC FxLMS}. In the GFANC–FxNLMS algorithm, a CNN in the co-processor predicts the weight vector for generating a control filter at the frame rate, while the FxNLMS algorithm continuously refines the generated filter at the sampling rate. The generated filter provides a good initialization for the FxNLMS algorithm. However, the dual filter update mechanism may lead to competitive updates, as small variations in the weight vector can repeatedly reinitialize the FxNLMS algorithm, interrupting its updating process and undermining system stability. To address this issue, an online clustering module is introduced to regulate weight vector updates, effectively avoiding unnecessary re-initializations and enhancing system stability. Simulation results validate the effectiveness of the online clustering strategy and show that GFANC–FxNLMS outperforms GFANC, FxNLMS, SFANC, and SFANC–FxNLMS in noise reduction performance.

\begin{figure}[tp]
\centering
\includegraphics[width=\linewidth]{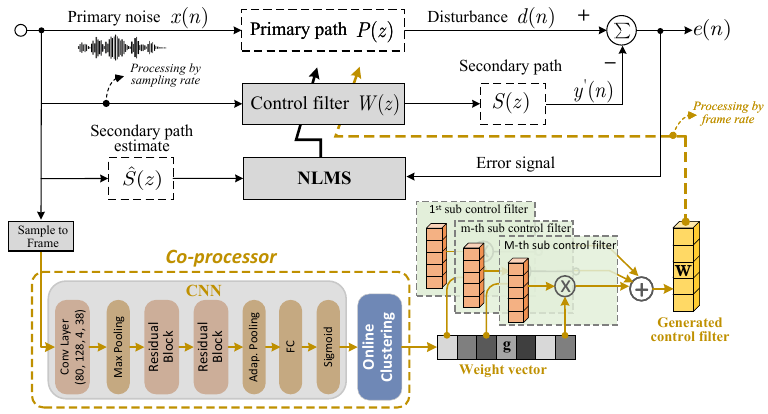}\vspace*{-0.4cm}
\caption{Block diagram of the GFANC–FxNLMS algorithm: a CNN and online clustering provide the weight vector to generate the control filter at the frame rate, while FxNLMS continuously optimizes the generated filter at the sampling rate.}
\label{Fig GFANC-FxNLMS}
\end{figure}

\vspace*{-0.4cm}
\section{The Hybrid GFANC-FxNLMS Algorithm with Online Clustering}
\begin{figure}[tp]
\centering
\includegraphics[width=\linewidth]{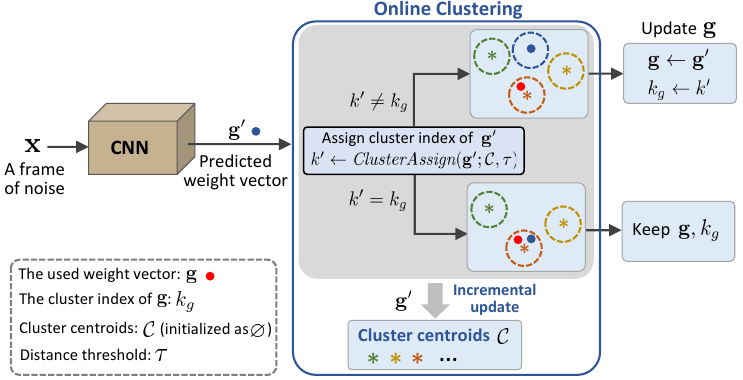}\vspace*{-0.3cm}
\caption{Online clustering decides if the CNN-predicted weight vector updates the current one, preventing unnecessary filter re-initializations in FxNLMS and enhancing stability.}
\label{Fig online clustering}
\end{figure}

\vspace*{-0.2cm}
The block diagram of the GFANC-FxNLMS algorithm is shown in Fig.~\ref{Fig GFANC-FxNLMS}. The method adopts a dual-rate architecture: at the frame rate, a CNN together with an online clustering module (Fig.~\ref{Fig online clustering}) provides the weight vector to generate the control filter, while at the sampling rate, the FxNLMS algorithm continuously optimizes the generated control filter.

\vspace*{-0.4cm}
\subsection{Sub Control Filters}
\begin{figure}[tp]
\centering
\includegraphics[width=0.43\linewidth, height=2.4cm]{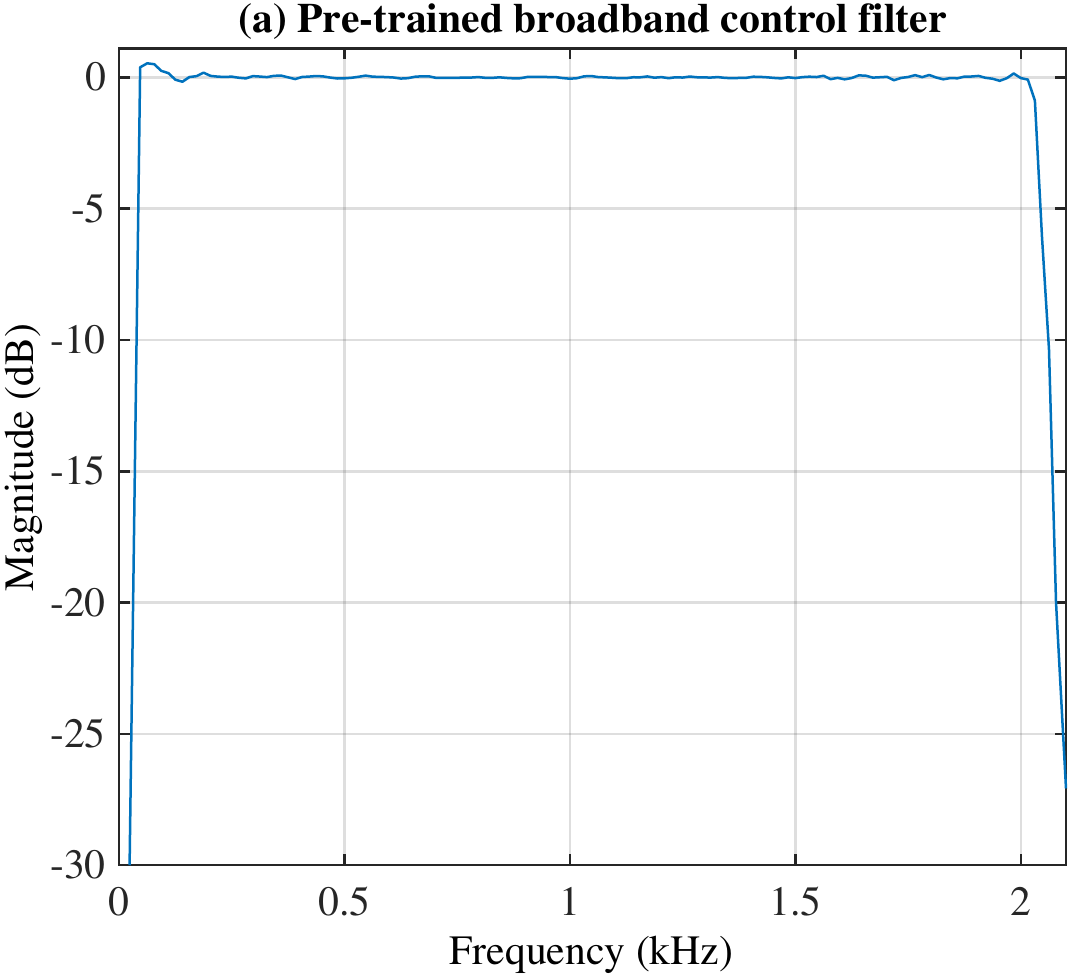}
\includegraphics[width=0.43\linewidth, height=2.4cm]{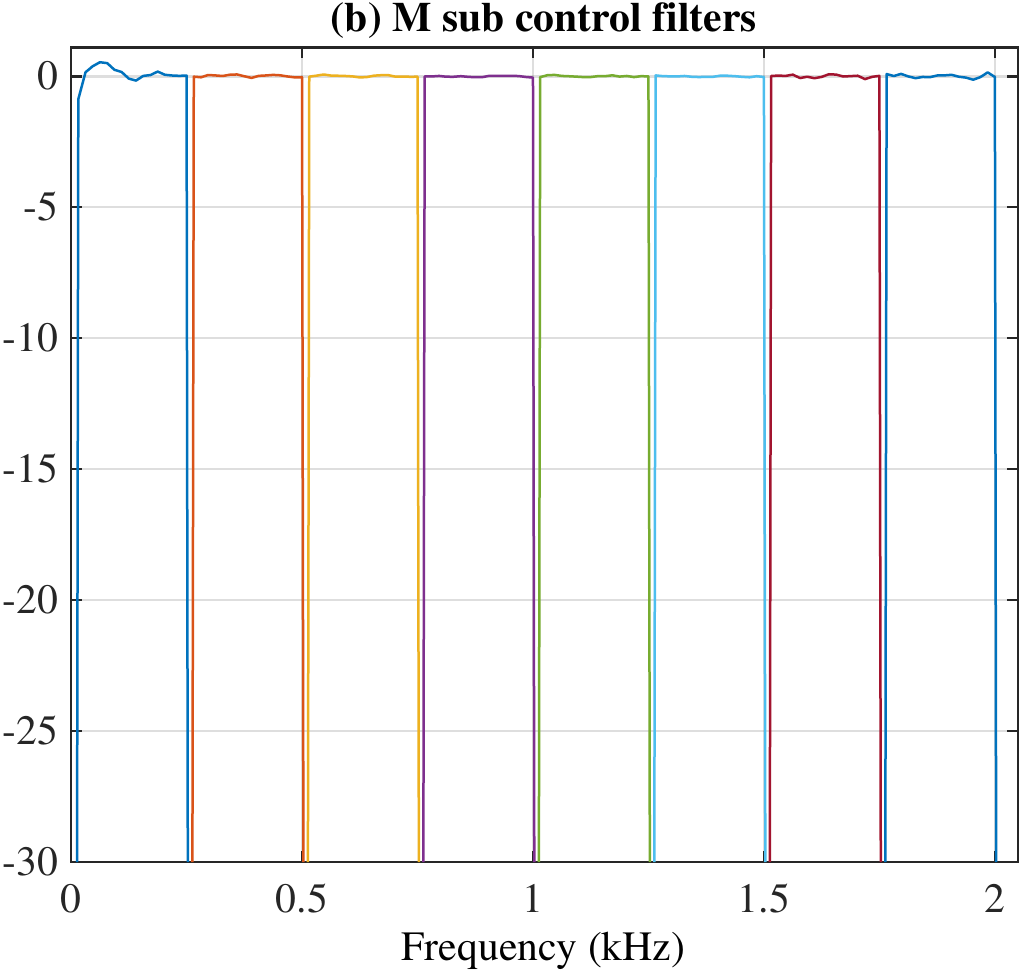}
\vspace*{-0.5cm}
\caption{Frequency spectra of a pre-trained broadband control filter and the $M$ sub control filters decomposed from it.}
\label{Fig Sub Control Filters}
\end{figure}

\vspace*{-0.2cm}
The construction of sub control filters is a crucial step in the GFANC–FxNLMS algorithm. First, the target ANC system is used to cancel a broadband primary noise covering the frequency components of interest. The optimal control filter is referred to as the pre-trained broadband control filter and serves as the sole prior information required by the GFANC–FxNLMS algorithm. Following the approach in \cite{LuoGFANCBayes}, the pre-trained broadband control filter is decomposed into $M$ sub control filters. The frequency spectra of the pre-trained broadband control filter and its sub control filters on synthetic acoustic paths are shown in Fig.~\ref{Fig Sub Control Filters}. The $M$ sub control filters form the sub control filter matrix $\mathbf{f}$, defined as
\begin{equation}
\setlength{\abovedisplayskip}{2pt}
\setlength{\belowdisplayskip}{2pt}
\mathbf{f} = \left[\mathbf{f}_1,\ldots,\mathbf{f}_m,\ldots,\mathbf{f}_M \right]^\mathrm{T},
\end{equation}
where $\mathbf{f}_m$ denotes the impulse response of the $m$-th sub control filter. By combining the sub control filters with different weights, control filters matched to the spectral characteristics of primary noises are generated.

\vspace*{-0.4cm}
\subsection{CNN Model}\vspace*{-0.1cm}
The CNN architecture used in the GFANC–FxNLMS algorithm is identical to that in the GFANC method \cite{LuoGFANCBayes}. It takes a frame of the noise waveform $\mathbf{x}$ as input and outputs a weight vector $\mathbf{g}'$ for combining the sub control filters, which belongs to a regression task. An adaptive labelling mechanism \cite{LuoGFANCBayes} is employed to automatically assign each noise frame its optimal weight vector. The elements of the weight vector take values in the range $0$ to $1$. When transferring the GFANC–FxNLMS method to a new acoustic environment, the trained CNN can be applied directly without retraining, and only the sub control filters need to be updated according to the acoustic paths.

\begin{table}[tp]
\centering
\caption{Pseudo-code of real-time noise cancellation in the hybrid GFANC–FxNLMS algorithm with online clustering.}
\resizebox{1.03\linewidth}{!}{
\renewcommand{\arraystretch}{1.05}
\begin{tabular}{|l|}
\hline
\textbf{Input:} Each frame of the reference signal is denoted as $\mathbf{x}$.\\
\textbf{Note:} The sub control filter matrix is $\mathbf{f}=\left[\mathbf{f}_1,\ldots,\mathbf{f}_M\right]^\mathrm{T}$. \\
The weight vector $\mathbf{g}=\left[g_{1},\ldots,g_{M}\right]$ is initialized as a zero vector. \\
$\textit{CNN}(\cdot)$ denotes the CNN parameterized by $\Theta^*$. \\
$k_{g}$ denotes the cluster index of $\mathbf{g}$ and is initialized as $0$.\\
\textbf{Online Clustering:} Given a set of centroids $\mathcal{C}$ (initialized as $\varnothing$) and\\ a threshold $\tau$, $\textit{ClusterAssign}(\mathbf{g}';\mathcal{C},\tau)$ outputs the cluster index $k'$ of $\mathbf{g}'$.\\[3pt]
\hline
\textbf{While} GFANC–FxNLMS on\\
~~~\textbf{\# Noise control and filter optimization (sampling rate):}\\
~~~\textbf{for} each sample of the reference signal \textbf{do}\\
~~~~~~$\mathbf{w}(n) = \mathbf{g}\,\mathbf{f} = \sum_{m=1}^{M} g_{m}\mathbf{f}_m$ \hfill $\triangleright$ Generate the control filter.\\
~~~~~~$e(n) = d(n) - s(n) * \!\left[\mathbf{w}^\mathrm{T}(n)\mathbf{x}(n)\right]$ \hfill $\triangleright$ Real-time noise control.\\
~~~~~~$\mathbf{w}(n\!+\!1)=\mathbf{w}(n)+\mu(n)\,\mathbf{x}'(n)\,e(n)$ \hfill $\triangleright$ FxNLMS updates the filter.\\
~~~\textbf{end for}\\[2pt]
~~~\textbf{\# Weight vector prediction in the co-processor (frame rate):}\\
~~~\textbf{for} each frame of the reference signal \textbf{do}\\
~~~~~~$\mathbf{g}' \leftarrow \textit{CNN}(\mathbf{x};\Theta^*), \;\mathbf{g}'=\left[g'_{1},\ldots,g'_{M}\right]$ \hfill $\triangleright$ Predict weight vector.\\
~~~~~~$k' \leftarrow \textit{ClusterAssign}(\mathbf{g}';\mathcal{C},\tau)$ \hfill $\triangleright$ Assign cluster index for $\mathbf{g}'$.\\
~~~~~~\textbf{if} $k' \neq k_{g}$ \textbf{then} \hfill $\triangleright$ Update when the cluster index changes.\\
~~~~~~~~~$(\mathbf{g}, k_{g}) \leftarrow (\mathbf{g}',\, k')$ \hfill $\triangleright$ Use $\mathbf{g}'$ and $k$ to update $\mathbf{g}$ and $k_{g}$.\\
~~~~~~\textbf{end if}\\
~~~\textbf{end for}\\
\hline
\end{tabular}
}
\label{Table GFANC-FxNLMS}
\end{table}

\vspace*{-0.3cm}
\subsection{Online Clustering Module}\vspace*{-0.2cm}
Minor variations in the weight vector, such as changing $[0.1, 0.5, 0.5, 0.5, 0.5, 0.5, 0.5, 0.5]$ to [0.2, 0.5, 0.5, 0.5, 0.5, 0.5, 0.5, 0.5], result in negligible differences in the generated control filters. Reinitializing the FxNLMS algorithm with marginally different control filters disrupts its adaptation process, resulting in competitive filter updates and compromising system stability. To address this problem, online clustering is employed to decide whether the CNN-predicted weight vector should replace the currently used one.

The online clustering is a dynamic process, as illustrated in Fig.~\ref{Fig online clustering}. Notably, the number of clusters $K$ is not fixed in advance, and centroids are updated incrementally with new $\mathbf{g}'$ \cite{OnlineClusteringPaper}. Each cluster is represented by a centroid, defined as the mean of all weight vectors assigned to the cluster. Given a set of centroids $\mathcal{C}=\{\mathbf{c}_1,\ldots,\mathbf{c}_K\}$ (initialized as $\varnothing$) and a distance threshold $\tau$, the cluster assignment operation is denoted as $\textit{ClusterAssign}(\mathbf{g}';\mathcal{C},\tau)$, which outputs the cluster index $k$ of $\mathbf{g}'$ according to
\begin{equation}
\setlength{\abovedisplayskip}{2pt}
\setlength{\belowdisplayskip}{2pt}
k' = 
\begin{cases}
K+1, & \text{if } \min\limits_{j\in\{1,\cdots,K\}}\|\mathbf{g}'-\mathbf{c}_j\|_2 > \tau, \\
\arg\min\limits_{j\in\{1,\cdots,K\}}\|\mathbf{g}'-\mathbf{c}_j\|_2, & \text{otherwise},
\end{cases}
\end{equation}
where $\mathbf{c}_j$ denotes the $j$-th cluster centroid, and $\|\cdot\|_2$ represents the Euclidean distance. If $k' = K+1$, $\mathbf{g}'$ lies outside all clusters and a new cluster is created. After each assignment, cluster centroids are updated incrementally as
\begin{equation}
\setlength{\abovedisplayskip}{2pt}
\setlength{\belowdisplayskip}{2pt}
\mathbf{c}_{k'} \leftarrow \frac{n_{k'} \mathbf{c}_{k'} + \mathbf{g}'}{n_{k'}+1},
\end{equation}
where $n_{k'}$ is the number of weight vectors previously assigned to the $k'$-th cluster.

The cluster index of the current weight vector $\mathbf{g}$ is denoted as $k_{g}$. Updating $\mathbf{g}$ with $\mathbf{g}'$ is performed only if the assigned cluster index $k'$ of $\mathbf{g}'$ differs from $k_{g}$:
\begin{equation}
\setlength{\abovedisplayskip}{2pt}
\setlength{\belowdisplayskip}{2pt}
(\mathbf{g}, k_{g}) \leftarrow (\mathbf{g}',\, k'), \quad \text{if } k' \neq k_{g}.
\end{equation}
The online clustering strategy is dynamic, as centroids are continuously updated with weight vectors. This allows the method to tolerate minor fluctuations while capturing significant changes, thereby preventing conflicting filter updates between GFANC and FxNLMS and enhancing system stability.

\begin{table}[tp]
\centering
\caption{Performance of the CNN on the testing dataset.}
\resizebox{0.9\linewidth}{!}{
\begin{tabular}{ll}
\hline
\textbf{Metric} & \textbf{Value} \\
\hline
Network input length & $16,000$ samples \\
MSE loss for predicting weight vector & $0.0031$ \\
Number of model parameters & $0.21$ million \\
Multiply-accumulate operations (MACs) & $237.56$ million \\
Floating-point operations (FLOPs) & $480.41$ million \\
\hline
\end{tabular}
}
\label{Table GFANC CNN Performance}
\end{table}

\begin{figure}[tp]
\centering
\includegraphics[width=0.495\linewidth, height=3cm]{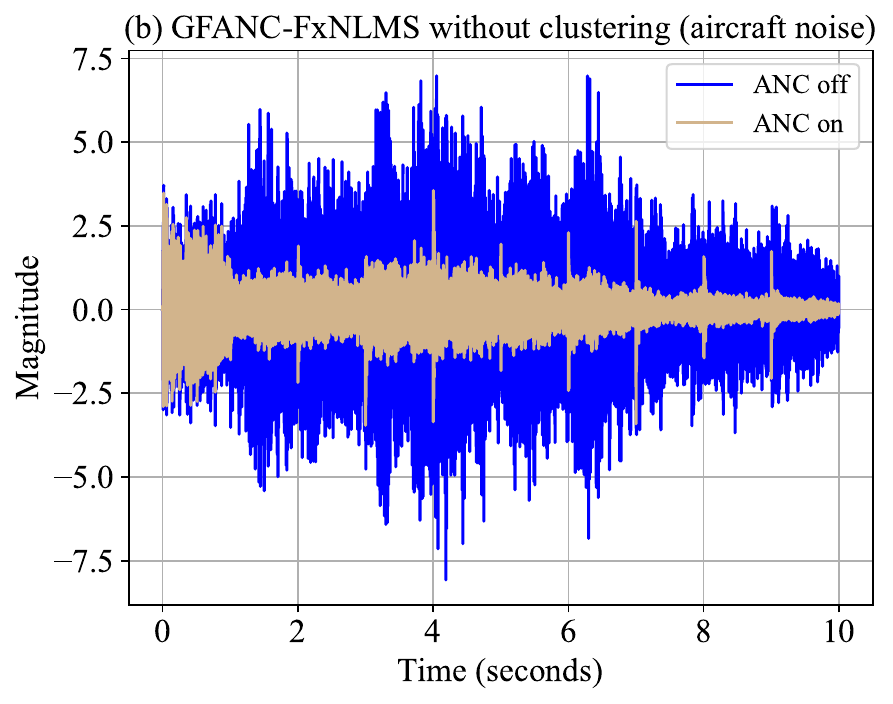}
\includegraphics[width=0.495\linewidth, height=3cm]{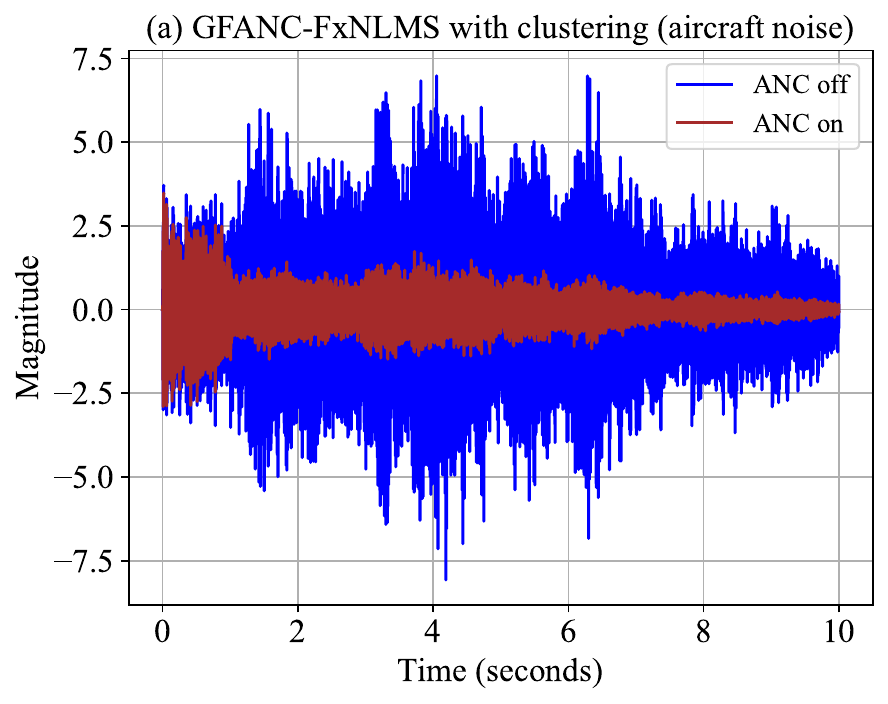}
\includegraphics[width=0.495\linewidth, height=3cm]{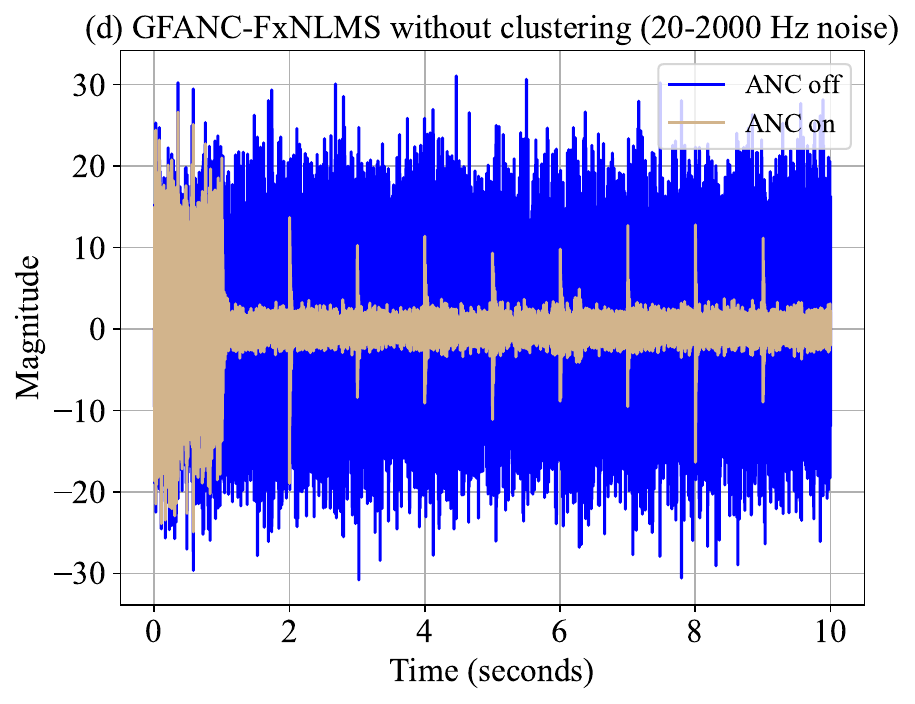}
\includegraphics[width=0.495\linewidth, height=3cm]{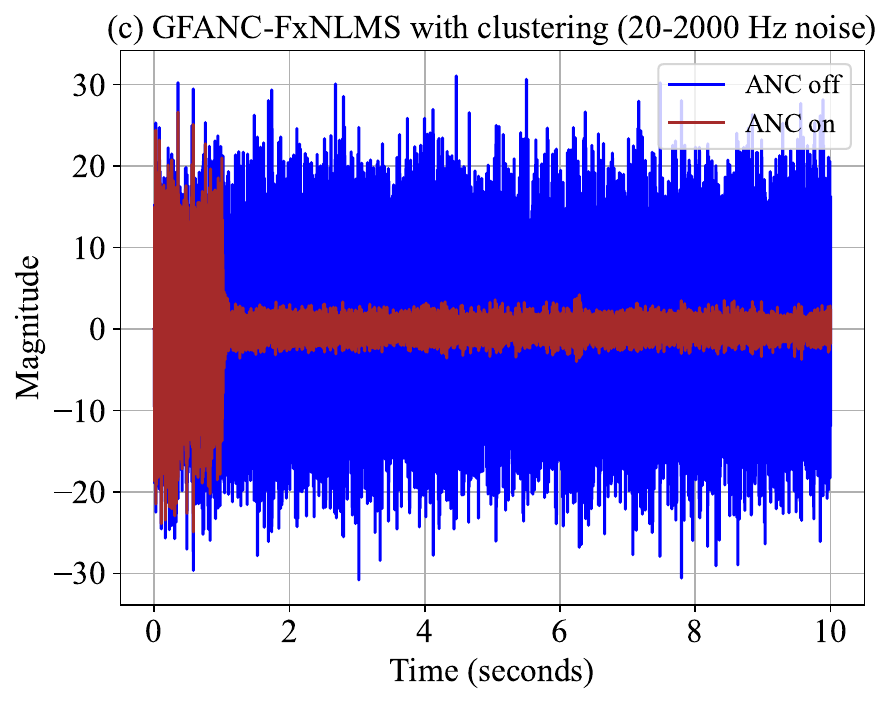}
\vspace*{-0.9cm}
\caption{Error signal comparison of the GFANC–FxNLMS algorithm without and with online clustering.}
\label{Fig Simulation Clustering}
\end{figure}

\begin{figure*}[tp]
\centering
\includegraphics[width=0.24\linewidth, height=3.2cm]{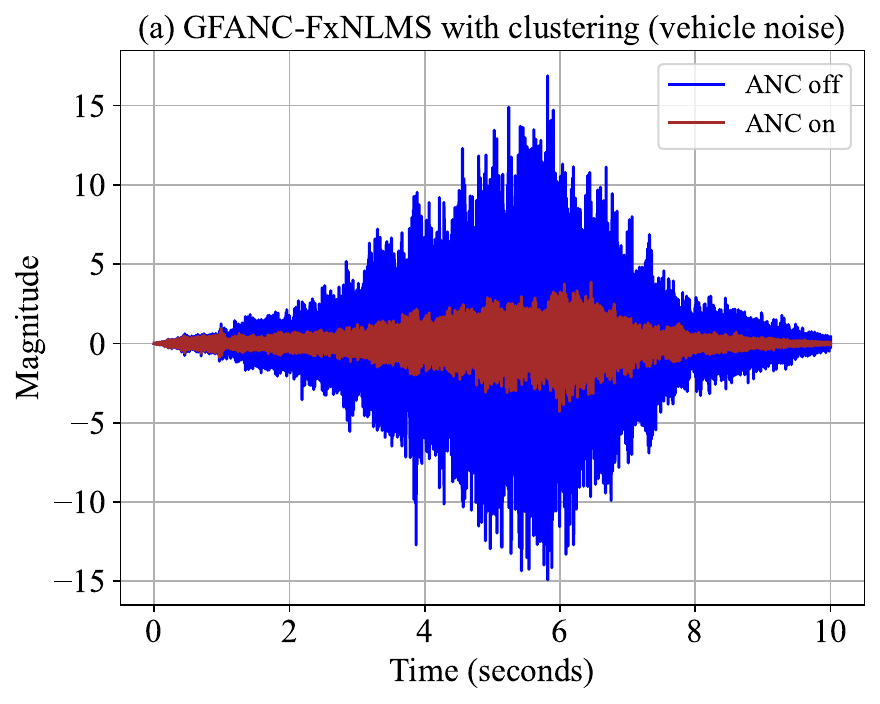}
\includegraphics[width=0.24\linewidth, height=3.2cm]{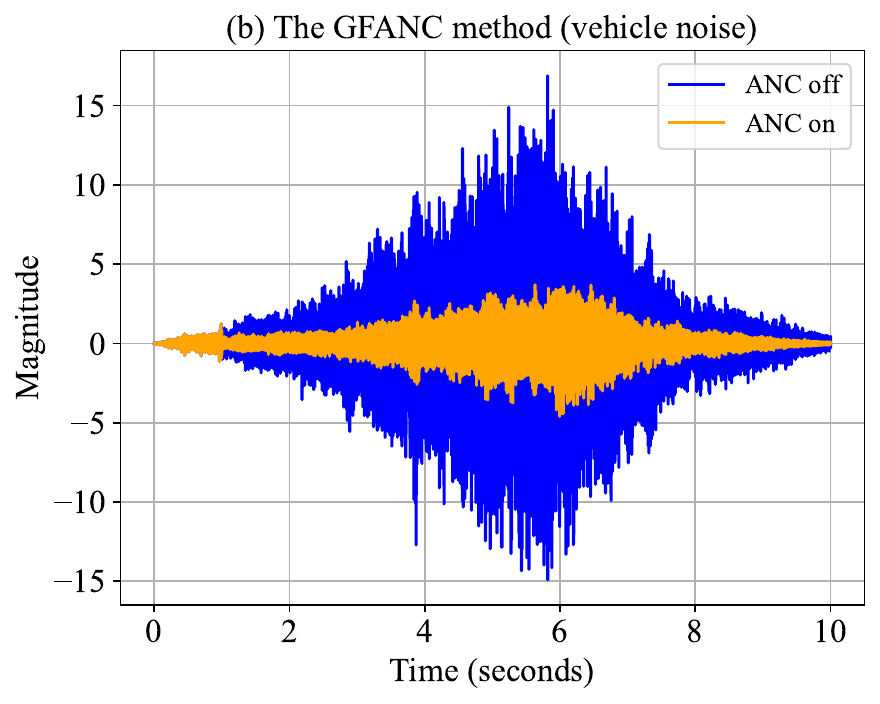}
\includegraphics[width=0.24\linewidth, height=3.2cm]{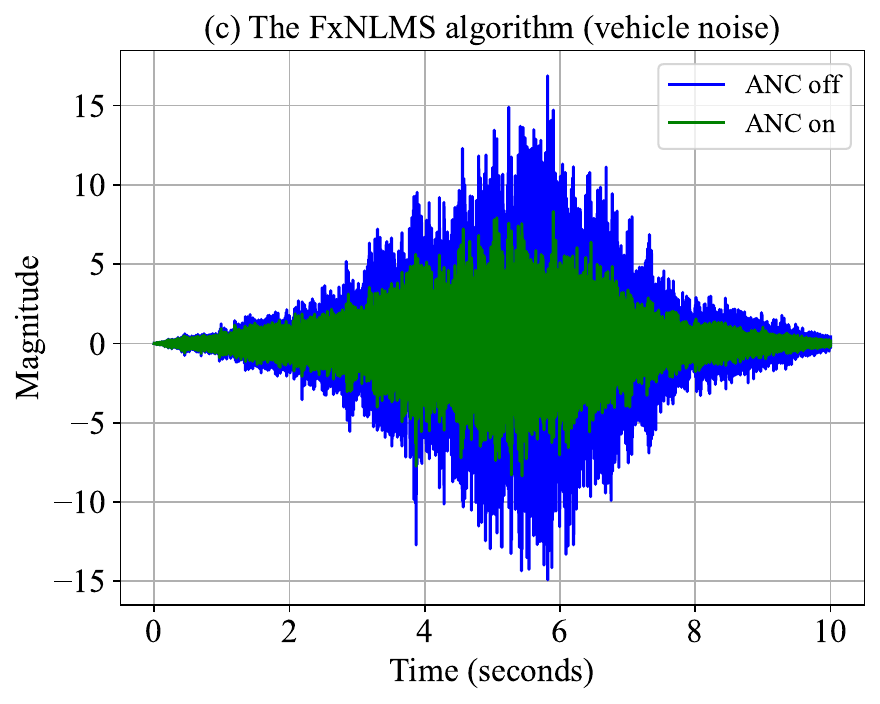}
\includegraphics[width=0.26\linewidth, height=3.2cm]{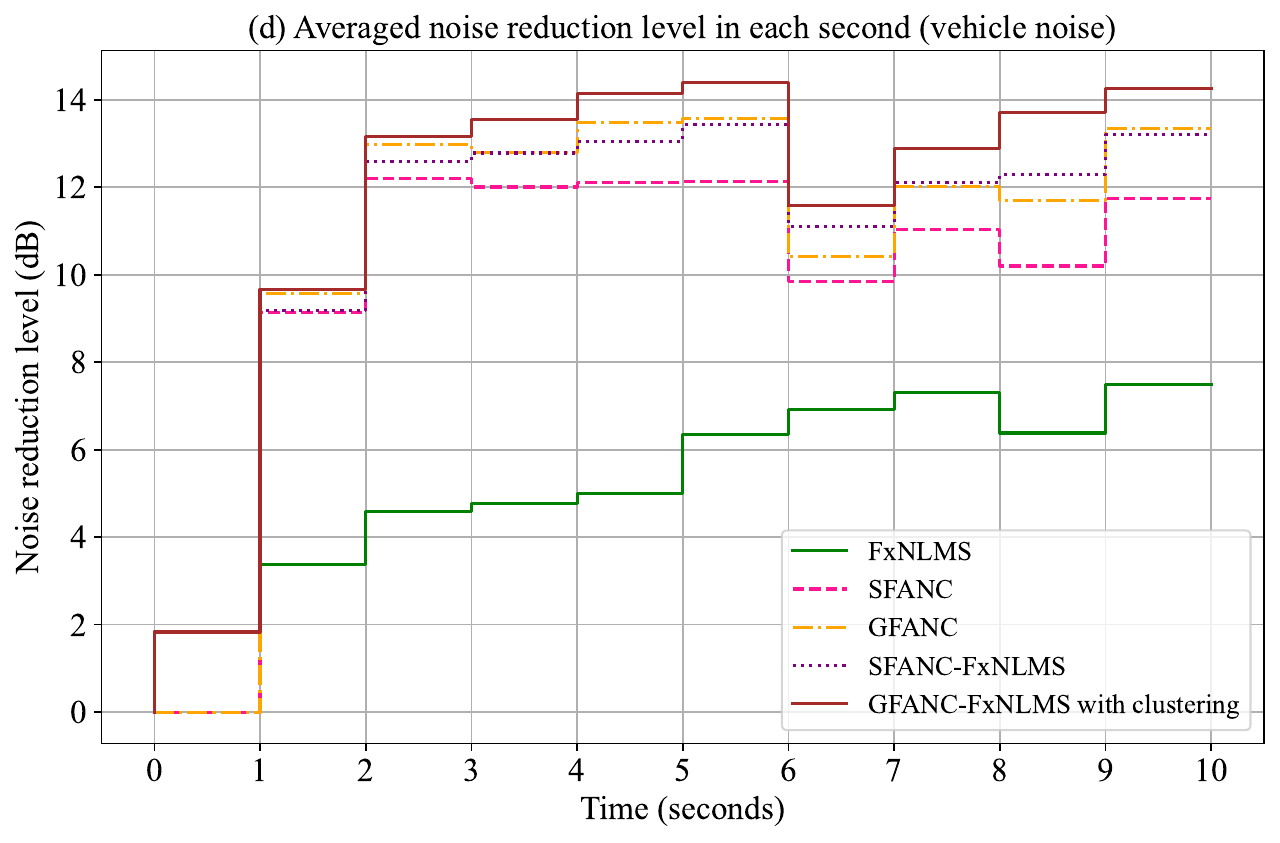}\\
\includegraphics[width=0.24\linewidth, height=3.2cm]{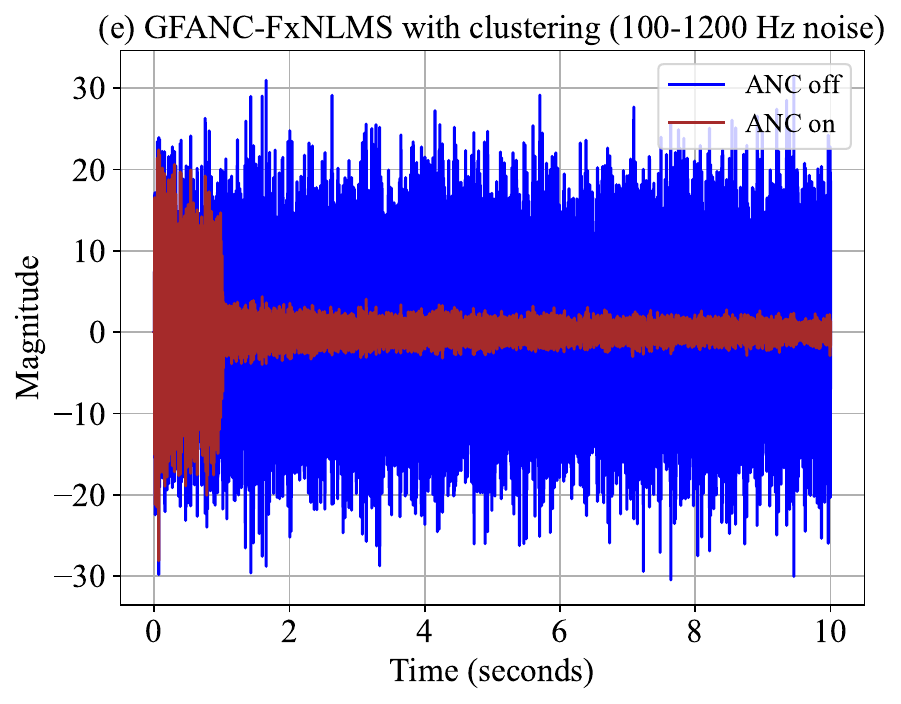}
\includegraphics[width=0.24\linewidth, height=3.2cm]{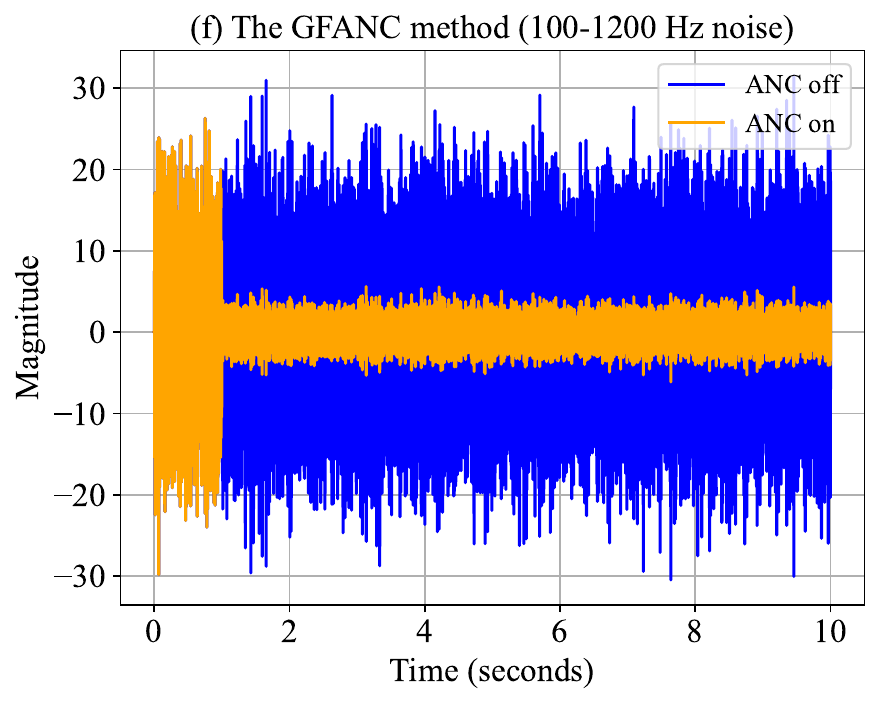}
\includegraphics[width=0.24\linewidth, height=3.2cm]{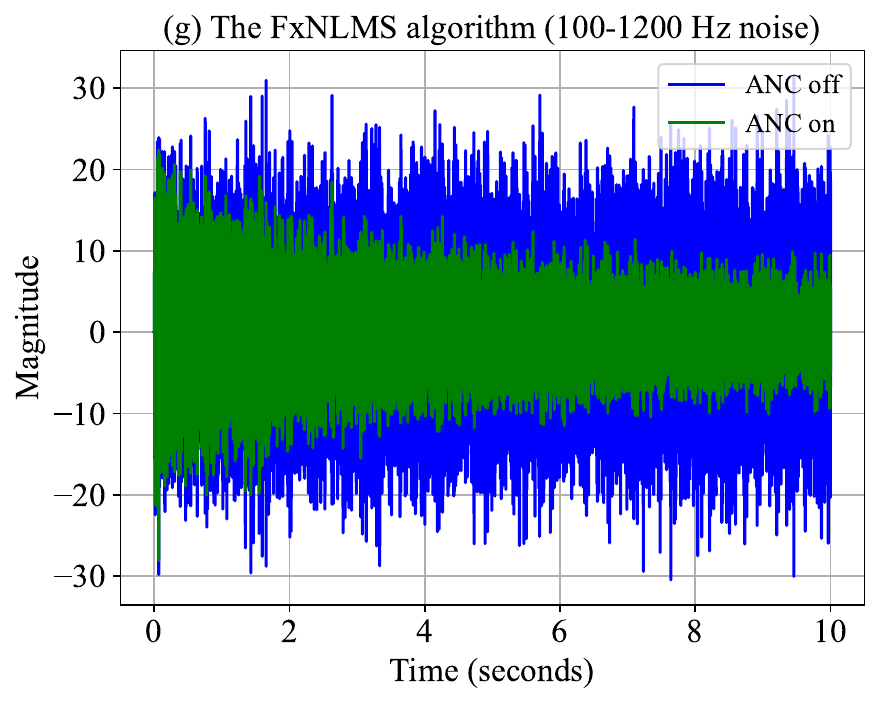}
\includegraphics[width=0.26\linewidth, height=3.2cm]{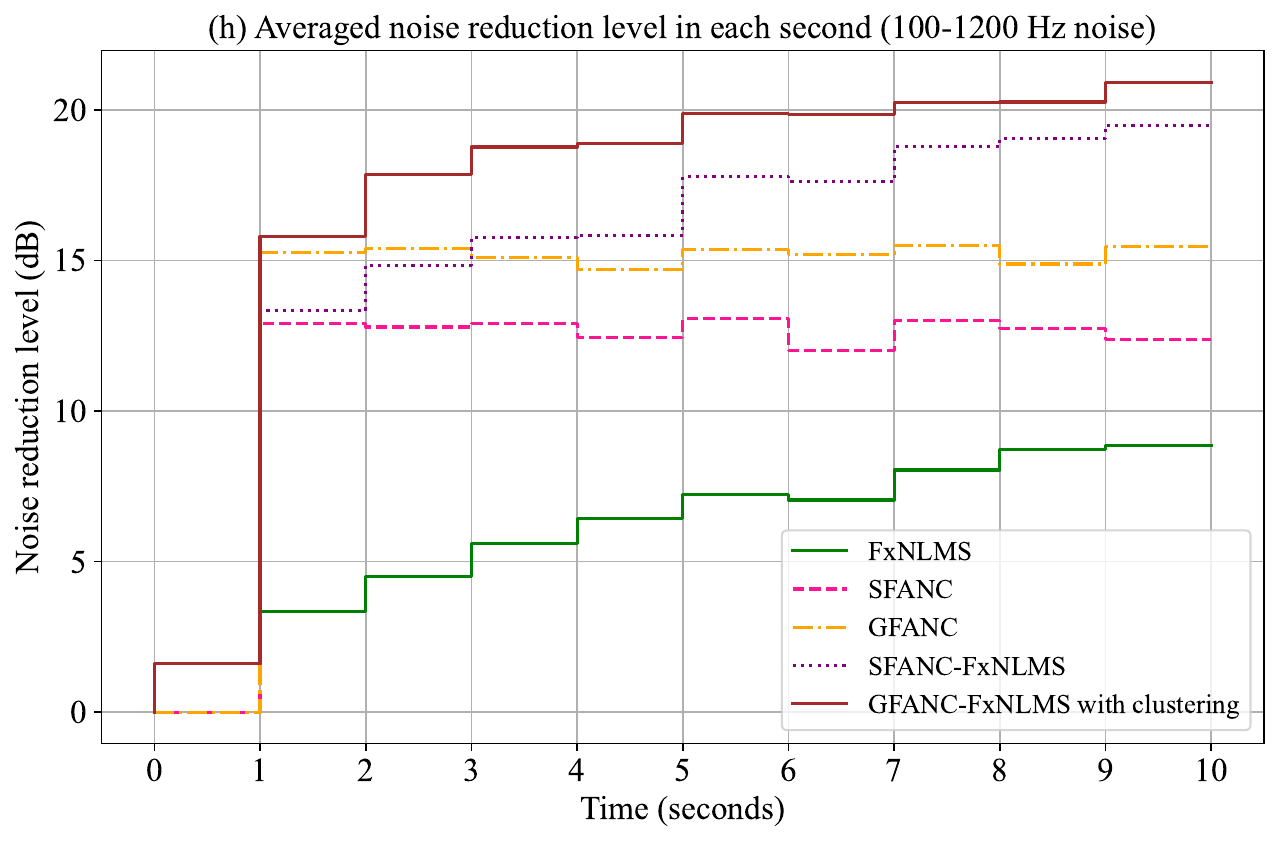}
\vspace*{-0.5cm}
\caption{Noise reduction performance of different ANC algorithms on (a)-(d): the vehicle noise, (e)-(h): the $100$-$1200$ Hz noise.}
\label{Fig Simulation Different ANC Methods}
\end{figure*}

\vspace*{-0.4cm}
\subsection{Real-time Noise Control}\vspace*{-0.2cm}
The real-time noise control process of the GFANC-FxNLMS algorithm is summarized in Table~\ref{Table GFANC-FxNLMS}, consisting of two parallel modules operating at different rates: (1) \textbf{Co-processor:} For each noise frame, the CNN predicts a weight vector $\mathbf{g}^\prime$. The online clustering module then determines whether $\mathbf{g}^\prime$ should update the current weight vector $\mathbf{g}$. (2) \textbf{Real-time controller:} $\mathbf{g}$ is used to generate a new control filter $\mathbf{w}$. Operating at the sampling rate, the controller performs noise cancellation and adaptively refines $\mathbf{w}$ using the FxNLMS algorithm. Efficient coordination between the co-processor and the real-time controller enables delayless noise control while using the feedback error signal to continuously optimize the control filter.

\vspace*{-0.6cm}
\section{Numerical Simulations}\vspace*{-0.2cm}
This section evaluates the effectiveness of the CNN and online clustering in GFANC–FxNLMS and compares its noise control performance with related ANC algorithms. The pre-trained broadband control filter covering the frequency range of $20$–$2,000$ Hz is decomposed into $8$ sub control filters. The control filter length and sampling rate are set to $1,024$ taps and $16$ kHz. $\tau$ is set to $0.6$. Real acoustic paths measured from the vent of a noise chamber are used.

\vspace*{-0.4cm}
\subsection{CNN Performance}\vspace*{-0.2cm}
For training the CNN, $80,000$ synthetic noises and $10,000$ real noises are used, while the testing dataset consists of $2,000$ synthetic noises and $500$ real noises. The real noises are taken from the \href{https://zenodo.org/records/3966543}{SONYC Urban Sound Tagging Dataset}. The synthetic noise frames are constructed by filtering white noise through various bandpass filters with random frequency ranges within $20$-$2,000$ Hz.

The performance of the CNN on the testing dataset is presented in Table~\ref{Table GFANC CNN Performance}. The Mean Squared Error (MSE) loss for predicting weight vectors is $0.0031$, showing that the CNN effectively extracts noise features to predict suitable weight vectors for different noises. The CNN model is also highly efficient, with only $0.21$ million parameters. With computational demands of $237.56$ million MACs and $480.41$ million FLOPs, the CNN exhibits very low complexity.

\vspace*{-0.4cm}
\subsection{Effectiveness of Online Clustering}\vspace*{-0.2cm}
To evaluate the effectiveness of the online clustering module, Fig.~\ref{Fig Simulation Clustering} compares the error signals of the GFANC–FxNLMS algorithm with and without online clustering. Under both aircraft noise and $20$–$2000$ Hz broadband noise, the algorithm with online clustering achieves a smoother noise control process and lower error signals. In contrast, without clustering, frequent frame-wise filter updates often reset the FxNLMS initialization, leading to unstable fluctuations. These results demonstrate that the online clustering module enhances stability by avoiding unnecessary re-initializations of FxNLMS.

\vspace*{-0.4cm}
\subsection{Comparison with Related ANC Methods}\vspace*{-0.1cm}
The GFANC–FxNLMS algorithm with online clustering is compared with several related ANC methods, namely GFANC, FxNLMS, SFANC, and SFANC–FxNLMS. The FxLMS algorithm uses a step size of $0.002$. The SFANC method employs $7$ pre-trained control filters with frequency ranges of \(20\)–\(2000\) Hz, \(20\)–\(1010\) Hz, \(1010\)–\(2000\) Hz, \(20\)–\(515\) Hz, \(515\)–\(1010\) Hz, \(1010\)–\(1505\) Hz, and \(1505\)–\(2000\) Hz. Noise reduction performances of the different ANC algorithms for vehicle noise and $100$–$1200$ Hz noise are presented in Fig.~\ref{Fig Simulation Different ANC Methods}.

The results in Fig.~\ref{Fig Simulation Different ANC Methods} show that the GFANC–FxNLMS algorithm achieves both high noise reduction levels and rapid response for the two noises. It outperforms the FxNLMS algorithm in response speed, as GFANC provides good filter initialization for adaptive updates. It also achieves lower steady-state errors than GFANC and SFANC, owing to the continuous optimization of FxNLMS. Moreover, in the early stage, the GFANC–FxNLMS algorithm delivers higher noise reduction levels than SFANC–FxNLMS, which requires additional time to catch up because GFANC supplies more suitable filters than SFANC.

\vspace*{-0.43cm}
\section{Conclusion}\vspace*{-0.3cm}
To effectively integrate the strengths of GFANC and FxNLMS, this paper proposes a hybrid GFANC–FxNLMS algorithm with online clustering. In this approach, the control filter generated by GFANC is continuously refined using FxNLMS with the feedback error signal. The GFANC–FxNLMS algorithm achieves both fast response and low steady-state errors, while requiring only a single pre-trained broadband filter. Furthermore, the online clustering module mitigates repeated filter reinitializations in the dual-rate GFANC–FxNLMS framework, thereby enhancing system stability. Simulation results demonstrate that the hybrid GFANC–FxNLMS algorithm achieves superior noise reduction performance compared with GFANC and FxNLMS individually, and also outperforms the SFANC and SFANC–FxNLMS methods.

\vfill\pagebreak
\bibliographystyle{IEEEbib}
\small
\bibliography{A}

\end{document}